\documentclass[a4paper,10pt,twocolumn,twoside]{article}

\usepackage{geometry}
\advance\oddsidemargin by -1in
\advance\evensidemargin by -1in
\advance\headsep by 2.8mm
\usepackage[numbers]{natbib}
\usepackage{authblk}

\usepackage{graphicx}
\usepackage{braket}
\usepackage{amsfonts, amssymb}
\usepackage{comment}
\usepackage{algorithm}
\usepackage{algorithmic}
\usepackage{tikz}
\usetikzlibrary{quantikz2}
\usepackage{url}
\usepackage{subcaption}
\usepackage{orcidlink}

\def\BibTeX{{\rm B\kern-.05em{\sc i\kern-.025em b}\kern-.08em
    T\kern-.1667em\lower.7ex\hbox{E}\kern-.125emX}}

\usepackage[normalem]{ulem}

\usepackage[compact]{titlesec}

\title{Parallel QEC Decoding Applied to\\ Distributed Quantum Computing}

\author[1]{Gabriele~Incardona}
\author[1]{Davide~Ferrari\orcidlink{0000-0002-4777-7234}}
\author[1,*]{Michele~Amoretti\orcidlink{0000-0002-6046-1904}}

\affil[1]{\small \textit{Quantum Software Laboratory}, Department of Engineering and Architecture, University of Parma, Parma, 43124 Italy (\href{https://www.qslab.unipr.it/}{https://www.qslab.unipr.it/})}
\affil[*]{\small Corresponding author: Michele Amoretti, michele.amoretti@unipr.it}

\date{}

\begin{document}

\maketitle

\begin{abstract}
\textbf{
A novel parallel approach is proposed for QEC decoding based on Belief Propagation with Ordered Statistics Decoding. The main idea is to pre-process the error vectors obtained from Belief Propagation by applying Singular Value Decomposition locally to sub-regions of the lattice. The proposed approach is applied to distributed quantum computers and evaluated in terms of complexity, accuracy, and scalability.
}
\end{abstract}


\maketitle

\section{Introduction}

Environmental decoherence is the process by which quantum systems lose their quantum properties due to interactions with the environment. 
Quantum Error correction (QEC) utilizes the idea of expanding the Hilbert space beyond what is needed to store a single qubit of information \cite{Terhal2015,Demartiiolius2024}. Logical qubits are formed by encoding quantum information across physical qubits. QEC is cyclically performed on the underlying physical qubits. Errors are detected without measuring the qubits directly; instead, entangled ancilla qubits are measured. Once detected, the errors are corrected. 

\textit{Surface codes}, which belong to the family of \textit{stabilizer codes}, stand as the most promising candidates for building near-term error-corrected qubits because of their two-dimensional architectures, the requirement of only local operations, and high tolerance to quantum noise \cite{Google2023,Google2025}. 

In this work, a novel parallel approach is proposed for QEC decoding based on Belief Propagation with Ordered Statistics Decoding (BP-OSD) \cite{Roffe2020}. The main idea is to pre-process the error vectors (LLR) obtained from Belief Propagation, applying Singular Value Decomposition (SVD) locally to sub-regions of the lattice. The proposed approach is applied to distributed quantum computers, i.e., systems composed of multiple quantum processing units (QPUs) \cite{Caleffi2024,Barral2025} connected by quantum links for sharing entangled states \cite{Amoretti2020}.

The paper is organized as follows. Section \ref{sec:background} introduces stabilizer codes in general and surface codes in particular. 
Section \ref{sec:related} discusses related works on parallel decoding techniques.
Section \ref{sec:opt-bposd} illustrates the proposed parallel approach to optimizing the BP-OSD decoder. Section \ref{sec:DSC} shows how to apply the optimized surface code to a distributed quantum computer. Section \ref{sec:simulations} presents and discusses simulation results. Finally, Section \ref{sec:conclusion} concludes the paper with an outline for future work.

\section{Background}
\label{sec:background}

The need for long coherence times is one of the most challenging issues that physicists and engineers face in their attempts to build quantum computers. In practice, because of quantum decoherence, qubits can be bit-flipped ($\ket{0} \leftrightarrow \ket{1}$) and phase-flipped ($\ket{0} \leftrightarrow \ket{0}$, $\ket{1} \leftrightarrow -\ket{1}$), not experiencing full flips but rather angular shifts of the qubit state by an angle \cite{Devitt2013}.

Current major efforts to build a quantum computer are based on surface codes \cite{Kitaev1997}, operated as stabilizer codes \cite{Gottesman1997}. In the following, the minimum background is provided to the reader.

\subsection{Stabilizer Codes}
\label{sec:stabilizer}

An $[[n,k,d]]$ stabilizer code encodes $k$ logical qubits into $n$ physical qubits. 
It is defined by $n-k$ independent stabilizer generators (denoted as \textit{checks}) forming an abelian subgroup $\mathcal{S}$ of the $n$-fold Pauli Group $\mathcal{G}_n$ (i.e., $S \subset \mathcal{G}_n$). $\mathcal{S}$ is denoted as the stabilizer set or stabilizer group. The parameter $d$ is the \textit{distance} of the code. It signifies the minimum number of physical qubit errors required to cause an undetectable logical error; the higher the number, the better.
Formally, $d$ is the minimum weight (number of non-identity Pauli operators) of an element in the normalizer $\mathcal{N}(\mathcal{S})$ of $\mathcal{S}$, but not in $\mathcal{S}$ itself. The normalizer $\mathcal{N}(\mathcal{S})$ is the set of operators that map $\mathcal{S}$ to itself via conjugation. The size of the normalizer is $4 \cdot 2^{n+k}$. In general, calculating $d$ is an NP-Hard problem. For small values of $n$ and $k$, calculating $d$ is feasible~\cite{Gottesman1997}.

By measuring the checks, one can compute the syndrome $\bar{s}$, a binary vector of length $n-k$ that describes the error that has occurred. The measurement must be done indirectly so that the codestate is not lost. Given a set of checks $\{M_1, M_2, ...,M_{n-k}\} \in \mathcal{S}$ and a Pauli error $E \in \mathcal{G}_n$, the $i^{th}$ element of the syndrome will capture the commutation relationships between the error and the $i^{th}$ check: $E M_i = (-1)^{s_i} M_i E$.
The process of measuring checks is called \textit{syndrome extraction}. The process of estimating the quantum error from the measured syndrome is called \textit{decoding}. Once a guess $E \in \mathcal{G}_n$ of the error is obtained by the decoder, the noisy quantum state will be successfully corrected by applying $E$ to the physical qubits (since the elements of $\mathcal{G}_n$ are self-inverse).

\subsection{Surface Codes}
\label{sec:surface}

In the original surface code by Kitaev \cite{Kitaev1997}, physical qubits (data qubits) and measurement qubits (checks) were arranged in a toroidal lattice. Later, planar codes were introduced, which are easier to implement. Here, the rotated planar code in a square lattice with a Calderbank-Shor-Steane (CSS) structure is considered, which is the most practical and relevant at the time of writing. With respect to traditional planar codes, the rotated ones use slightly fewer qubits to correct the same number of errors.

\begin{figure}
\centering
\includegraphics[width=0.9\linewidth]{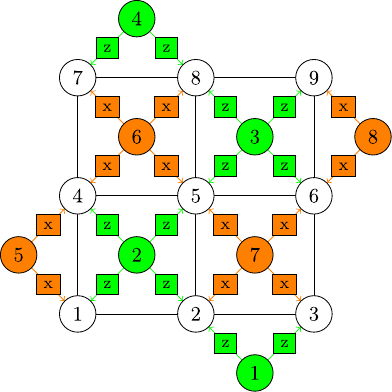}
\caption{Distance-3 rotated planar code \cite{Demartiiolius2024}.} 
\label{fig:d3rotated}
\end{figure}

In CSS surface codes, there are two types of checks: $X$-checks and $Z$-checks, which detect $X$-errors and $Z$-errors, respectively. An example is provided in Fig. \ref{fig:d3rotated}, showing a distance-3 rotated planar code. Data qubits are represented as white circles, $X$-check qubits as green circles, and $Z$-check qubits as orange circles. The stabilizing operators between the check qubits and the adjacent data qubits are represented as squares.
In the distance-3 planar code of Fig. \ref{fig:d3rotated}, $n = 9$ and $k=1$. Therefore, there are $n-k = 8$ checks: $Z_2Z_3$, $Z_1Z_2Z_4Z_5$, etc.

\begin{figure}[ht!]
\centering
\begin{subfigure}[b]{0.9\linewidth}
    \centering
    \includegraphics[width=\textwidth]{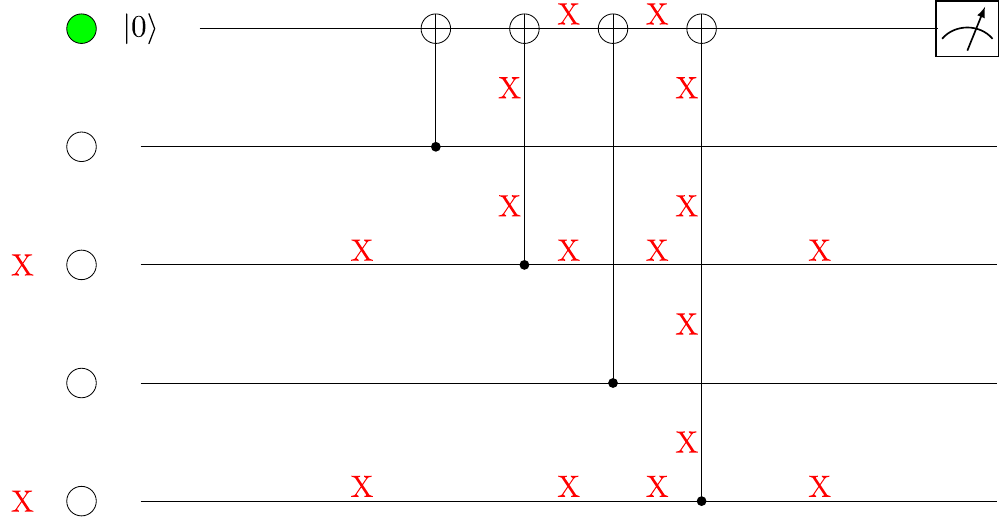}
    \caption{$X$-check qubits circuit.}
    \label{fig:check-circuit-x}
\end{subfigure}
\hfill
\begin{subfigure}[b]{0.9\linewidth}
    \centering
    \includegraphics[width=\textwidth]{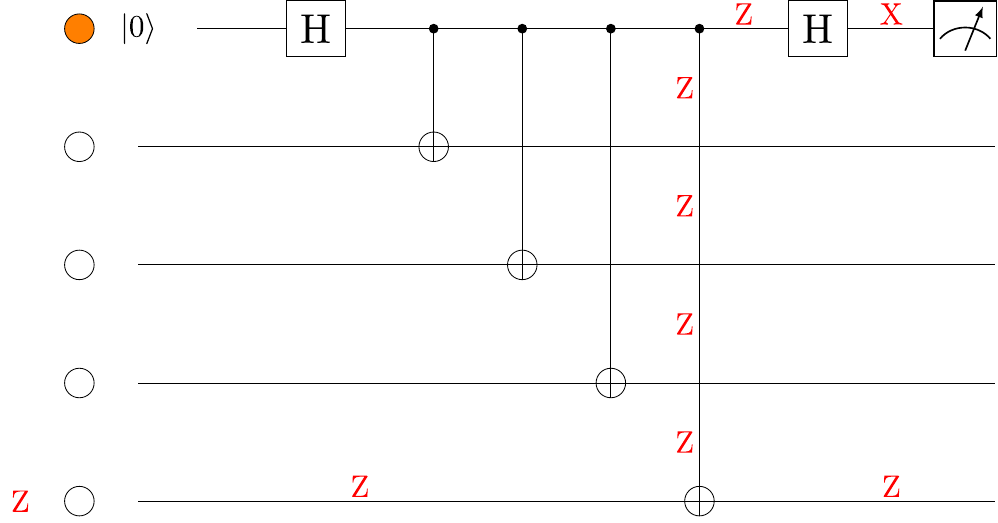}
    \caption{$Z$-check qubits circuit.}
    \label{fig:check-circuit-z}
\end{subfigure}
\caption{Stabilizing circuits for the check qubits. Error propagation examples are shown \cite{Demartiiolius2024}.}
\label{fig:check-circuits}
\end{figure}

Figure \ref{fig:check-circuits} illustrates the circuits for performing the $X$-checks and $Z$-checks. Check qubits are always initialized to $\ket{0}$. If an odd number of adjacent data qubits are affected by an $X$ or $Z$ error, this is detected by the measurement of the corresponding check qubit. Conversely, an even number of errors cannot be detected.
The circuits for performing the $X$-checks and $Z$-checks must be executed in a loop for the entire duration of the quantum computation with the logical qubits. As soon as the state of the data qubits is projected into the codespace, the measurement outcome becomes deterministic (if no error is applied).

\subsection{Belief Propagation based Decoder}
\label{sec:bpd}
Belief Propagation (BP) \cite{Demartiiolius2024} is a probabilistic decoding algorithm that estimates the error probability of each individual qubit based on the information provided by parity checks.
The algorithm operates on a representation of the code called a Factor Graph. This is a bipartite graph consisting of Variable Nodes, which represent the data qubits, and Factor Nodes, which represent the stabilizers (parity constraints based on the measured syndrome).

At the heart of BP is the iterative exchange of messages between nodes along the graph’s edges. 
The qubit communicates to the check its “estimate” of whether it should be flipped based on data received from the other checks. The check communicates to the qubit the probability that it must be flipped to satisfy the measured syndrome, based on information received from the other qubits connected to it.
At each iteration, the estimate of the error probability is refined. The algorithm terminates when the estimated syndrome converges with the measured one or after a maximum number of steps. BP can easily incorporate detailed information about physical noise. The advantage is that it is extremely fast and parallelizable, making it suitable for rapid decoding.

Despite its efficiency, BP faces two main obstacles in the surface code: 
    \begin{itemize}
        \item Short Loops: The Surface Code lattice contains many loops (four qubits per stabilizer). These create feedback loops where messages “bounce” and amplify themselves, preventing convergence.
        \item Degeneracy: When many different error chains lead to the same syndrome, BP can remain “undecided” between several equivalent solutions, leading to logical errors.
    \end{itemize} 
To overcome these limitations, current research often combines BP with the Ordered Statistics Decoding (OSD) algorithm \cite{Roffe2020}.
The process consists of two phases: 
\begin{itemize}
\item Algebraic post-processing: Once the BP has completed its iterations, it provides a probability vector called \textit{Log-Likelihood Ratios} (LLR) for each qubit. The OSD uses this information to rank the qubits from most reliable to least reliable.
\item Solving the linear system: The OSD selects a subset of qubits (a basis for the parity matrix $H$) based on their statistical reliability and solves the syndrome equation $H \cdot e = s$ using Gaussian elimination.
\end{itemize}
This combined approach makes it possible to “force” a valid solution that satisfies the measured constraint, leveraging the speed of the BP algorithm to drastically reduce the search space of the OSD.

BP-OSD exhibits a worst-case time complexity of $O(n^3)$; a bottleneck stems specifically from the OSD post-processing stage, whereas the initial BP stage is significantly lighter.

\section{Related Work}
\label{sec:related}
In recent years, the development of high-rate Quantum Low Density Parity Check (QLDPC) codes has shifted the research focus towards the challenge of scalability. These codes often result in an increase in decoding latency due to the complexity of the parity-check matrices. Consequently, several articles have been published dedicated to the optimization of decoding algorithms, specifically BP-OSD variants. In this section, three recent and relevant works are analyzed.

Hillmann et al.\cite{Hillmann2024} outlined the computational bottleneck of the OSD stage, particularly for large-scale codes. To address this, they proposed a lightweight alternative designed to achieve accuracy comparable to or even exceeding the standard BP-OSD model while significantly reducing decoding latency. Nevertheless, this algorithm is limited by the requirement for a central controller to manage an exceptionally high number of qubits. Furthermore, parallelization can only be initiated after error detection has occurred. Processing such large parity-check matrices on a single centralized decoder poses significant scalability challenges as the system size grows.

The decoding strategy proposed by Wang et al.~\cite{Wang2026} introduces a fully parallelizable post-processing framework designed to improve the performance of BP decoders for Quantum LDPC (QLDPC) codes. Unlike BP-OSD approaches, the proposed framework avoids the Gaussian elimination stage required by OSD, thereby significantly reducing computational complexity and decoding latency. However, the effectiveness of the method depends on several heuristic parameters. Consequently, the performance of the decoder may vary depending on the underlying code family and the adopted noise model, requiring parameter tuning to achieve optimal performance.

The work proposed by Fan et al.~\cite{Fan2026} introduces a lightweight pre-processing technique aimed at improving the efficiency of BP-based decoders. The proposed method analyzes local syndrome patterns obtained from circuit-level syndrome measurements in order to identify error configurations that are likely to correspond to elementary fault events. By guiding the BP process with syndrome-informed prior information, the proposed approach reduces the number of BP iterations required for convergence while preserving, and in some cases improving, the overall logical error rate performance.
However, although the proposed method improves the convergence behavior of the BP phase, the OSD stage may still be required in high physical error-rate regimes, where BP alone is more likely to fail to converge to a valid correction.

\section{BP-OSD Optimization via Local SVD}
\label{sec:opt-bposd}

Since the Ordered Statistics Decoding (OSD) phase has a computational complexity of $O(n^3)$, it constitutes the main bottleneck for the scalability of the decoder in large-distance codes. The proposed approach introduces pre-processing of the error vectors (LLR) obtained from Belief Propagation, applying the Singular Value Decomposition (SVD) locally on sub-regions of the lattice.

In this architecture, the code is divided into $M$ blocks, each containing $m$ qubits. SVD is run in parallel on each block to identify the dominant error vectors, i.e., the principal components of the local probability distribution, which are then transmitted to a central coordinator. The coordinator assembles the reduced-order matrices to compute the final corrections via the OSD. From a complexity standpoint:

\begin{itemize}
    \item Each local computational unit performs local SVD by processing a small matrix. The complexity is $O(m^3)$, where $m \ll n$.
    \item The global OSD algorithm no longer has to operate on the entire $n$ qubit space but on a reduced basis of dominating vectors.
    \item If $r$ is defined as the compression factor obtained via SVD (keeping only singular values above a certain threshold $\epsilon$), the complexity of SVD is $O((n/r)^3)$. 
    \item Overall, the global OSD has a complexity of $O(m^3 + (n/r)^3)$. In the worst case, $r=1$, so the OSD complexity is $O(n^3)$ as in the non-parallel case. In the best case, $r=m$, so the OSD complexity becomes $O(m^3 + M^3)$, which—regardless of the choice of $m$ and $M$—is always less than the non-parallel complexity.
\end{itemize}

This method allows us to shift the computational load to a massively parallelizable stage (local SVD), drastically reducing the size of the linear system that the OSD must solve globally, without sacrificing the statistical accuracy guaranteed by the probabilistic model.

\section{Distribution of a Surface Code}
\label{sec:DSC}

Fault-tolerant distributed quantum computing (FT-DQC) can be approached in different ways~\cite{Chandra2025}: connecting small modules and using GHZ states to perform stabilizer checks (Type-I); distributing large error correcting codes across multiple physical modules (Type-II) by means of non-local CNOT gates; or allowing each node to operate an entire logical code block (Type-III), with fault-tolerant computations between nodes enabled through non-local operations such as transversal gates, teleportation of logical gates, or distributed lattice surgery procedures.

\begin{figure}
\centering
\includegraphics[width=0.9\linewidth]{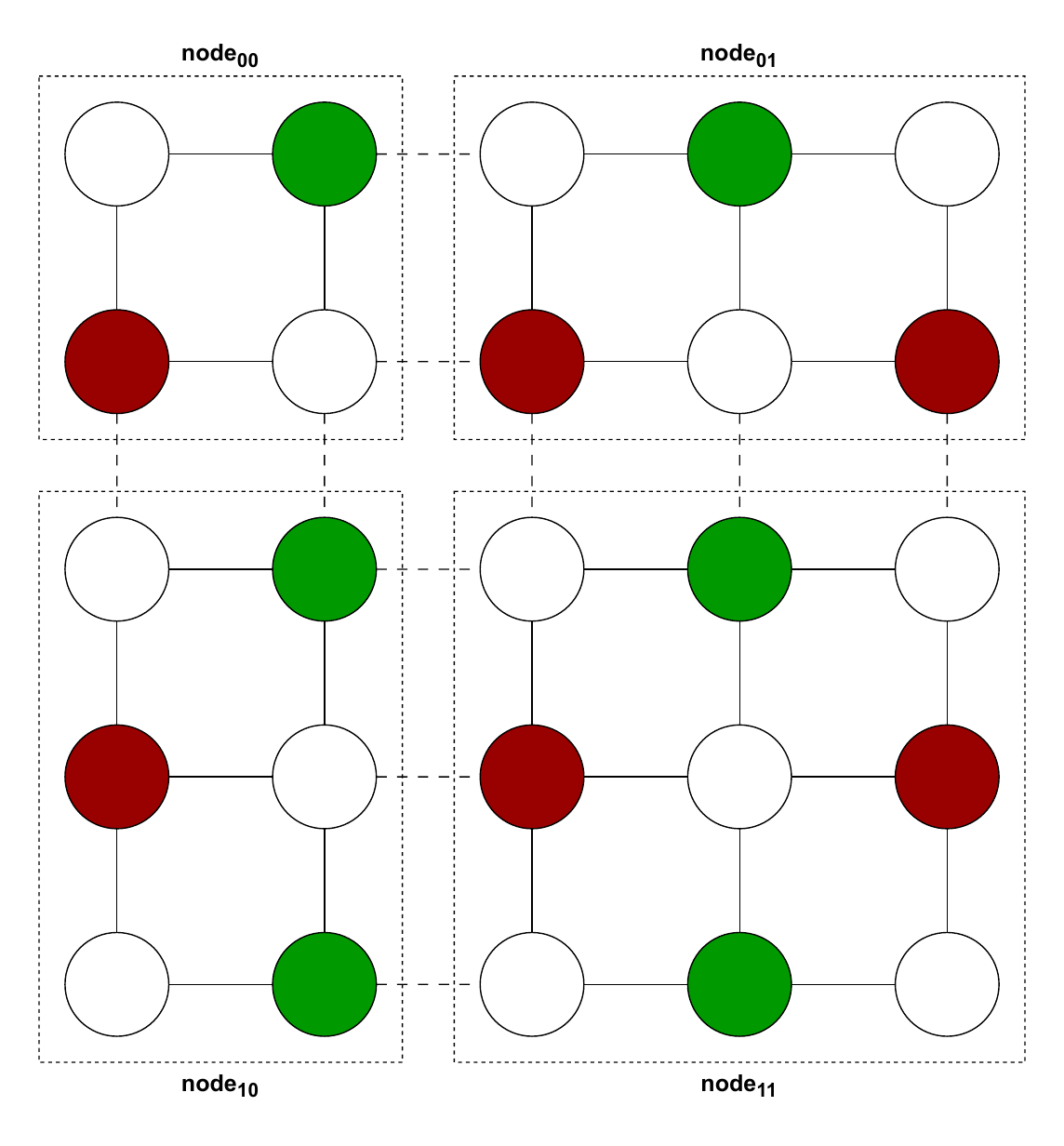}
\caption{Representation of a $5 \times 5$ surface code distributed across 4 QPUs. White circles are data qubits, red circles are Z-stabilizers and green circles are X-stabilizers. Solid line connections denote local interactions within the QPUs, while dashed lines denote non-local CNOT operations.}
\label{fig:surface_code}
\end{figure}

In this work, the Type-II approach is adopted.
As illustrated in Fig. \ref{fig:surface_code}, the surface code is extended to DQC architectures by dividing the planar lattice into $M$ computational nodes, each containing $m$ qubits. Code continuity along the node boundaries is ensured by edge qubits that can be entangled (\textit{ebit}). In this configuration, the syndrome computation occurs in two modes:
\begin{itemize}
\item \textbf{Internal:} Local stabilizers are measured directly inside the node.
\item \textbf{Distributed:} For boundary stabilizers, parity information is extracted via quantum communication protocols based on entanglement and subsequent classical bit transmission. 
\end{itemize}

The distributed version of the surface code fits well with the solution for optimizing BP-OSD via local SVD presented in the previous section.

\section{Simulations}
\label{sec:simulations}

The functional and performance evaluation of the proposed protocol was carried out using \textit{SquidASM} \cite{SquidASM}, an advanced software framework designed to model and simulate quantum network applications.
SquidASM allowed for a granular investigation of quantum noise in terms of various error channels, including Identity, Hadamard, Initialization, Readout, and CNOT gate errors. Additionally, an "All-Errors" configuration was implemented to simulate a comprehensive noise environment in which all the aforementioned error types are injected simultaneously, providing a more realistic stress test for the distributed surface code.

\begin{figure}
\centering
\includegraphics[width=0.9\linewidth]{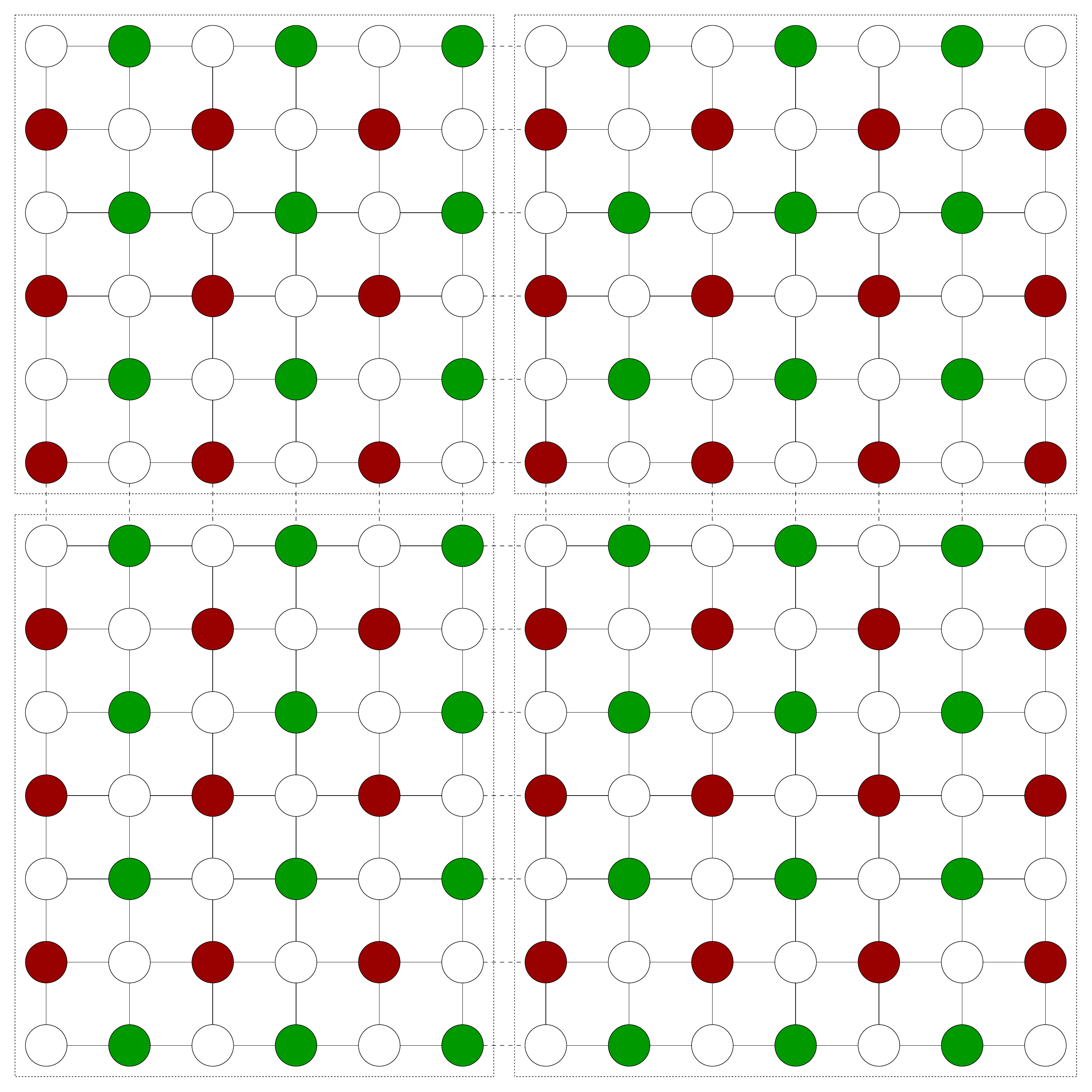}
\caption{The $13 \times 13$ qubit used for most simulations.}
\label{fig:simulation}
\end{figure}

The physical error probability $p_{\text{err}}$ was varied to analyze the protocol response to different noise intensities. The study focused on a $13 \times 13$ qubit lattice (distance $d=7$ for the surface code) distributed over 4 nodes (Fig. \ref{fig:simulation}). This compact network served as a fundamental proof-of-concept, allowing for a detailed analysis of distributed parity checks and inter-node synchronization within a controlled and computationally feasible environment.

The study was conducted through a comparative performance analysis consisting of two distinct simulation configurations. This approach is designed to evaluate the impact of the reduction in dimensionality on the decoding process.

In the \textbf{Baseline Configuration}, the SVD energy threshold is set to its maximum value of 1.0. This threshold represents the minimum amount of information preserved during compression, meaning that in this configuration, no singular values are truncated. This ensures that no information is discarded, allowing the BP-OSD algorithm to process the original, full-rank parity-check matrix. This configuration serves as a benchmark for standard decoding performance.

In the \textbf{SVD-Optimized Configuration}, an energy threshold of 0.98 is manually enforced. In this case, the SVD retains only the principal components that represent 98\% of the matrix energy, effectively discarding the remaining 2\% of less significant information. This results in a compressed parity-check matrix, allowing for an assessment of the trade-off between reduced computational complexity and logical error rate.

By executing these two scenarios independently, it is possible to isolate the effects of SVD-based matrix reduction on the decoder's accuracy and efficiency.

\subsection{Accuracy} 
\label{sec:accuracy}

The results presented in Table \ref{tab:simulation_result} illustrate the decoder's accuracy, defined as the number of successful error corrections achieved over 1000 independent simulation trials.

\begin{table}
    \centering
    \scriptsize 
    \caption{Comparison of Logical Accuracy with and without SVD optimization.}
    \label{tab:simulation_result}
    \begin{tabular}{|l|c|c|c|}
        \hline
        \textbf{Error Type} & \textbf{$p_{\text{err} [\%]}$} & \textbf{Baseline SVD[\%]} & \textbf{SVD-Optimized [\%]}  \\ \hline
        Identity      & 1.0      & 98.2                             & 99.5                     \\ \hline
        Identity     & 10.0        & 64.0                              & 64.9                   \\ \hline
        Hadamard      & 1.0       & 97.4                              & 99.0                    \\ \hline
        Hadamard      & 10.0       & 57.4                             & 58.6                    \\ \hline
        Initialization    & 1.0          & 91.6                              & 92.8                   \\ \hline
        Initialization    & 10.0         & 66.3                             & 67.5                    \\ \hline
        Readout         & 1.0   & 92.8                              & 94.1                    \\ \hline
        Readout        & 10.0     & 66.6                              & 64.7                  \\ \hline
        CNOT        & 1.0    & 82.1                              & 84.0                    \\ \hline
        CNOT       & 10.0      & 50.1                              & 52.3                    \\ \hline
        All        & 1.0    & 71.9                              & 74.4                     \\ \hline
        All        & 10.0     & 51.6                              & 50.0                    \\ \hline
    \end{tabular}
\end{table}

As shown in Table \ref{tab:simulation_result}, the SVD-optimized model generally outperforms the standard model, particularly at lower error rates. This improvement stems from the SVD's ability to isolate the main components of the signal, effectively filtering out stochastic noise. By treating certain perturbations as negligible singular values, the system avoids processing non-essential noise and focuses on correcting errors that significantly impact the logical outcome. 

At an error probability $p_{\text{err}} = 1.0\%$ -- which represents a challenging yet realistic worst-case scenario for individual gate operations -- the model maintains high logical accuracy. However, a significant performance drop is observed in the \textit{"All"} category. This is primarily because the errors are treated as independent variables in the simulation; consequently, when multiple error types are active simultaneously, the cumulative physical error rate per qubit increases drastically. Given a code distance of $d = 7$, the system can correct up to $t = 3$ errors. In the \textit{"All"} scenario, the high density of faults likely exceeds this threshold, overwhelming the model's corrective capacity.

The simulation with $p_{\text{err}} = 10\%$ serves as a stress test to observe the performance degradation under extreme, albeit unrealistic, conditions. It is noteworthy that while the decoder generally remains more accurate than random chance (50\%), the SVD optimization can become counterproductive at this threshold.

A crucial observation from these simulations is the impact of SVD optimization on computational efficiency. The parity-check matrix $H$, once processed via SVD, produces a reduced matrix ($H_{reduced}$) with an average compression factor $r = 2.17$. By reducing the effective number of qubits to less than half, the system achieves a significant reduction in decoding complexity. Since the complexity of Ordered Statistics Decoding (OSD) scales cubically with the number of variables, this dimensionality reduction translates to a speedup of at least eight times ($2^3$) compared to the non-optimized version, making the model substantially more scalable for real-time error correction.

\subsection{Global SVD in Distributed Architectures}
\label{sec:global}
A comparative analysis was conducted for a distributed surface code scenario where SVD optimization is performed globally by the coordinator. Following the configuration outlined in Section \ref{sec:accuracy}, local parity-check submatrices are computed at each QPU and transmitted to a central coordinator, which reconstructs the global matrix $H$. The SVD algorithm is then applied to this aggregated structure to compress the matrix prior to the OSD decoding phase. While this approach is satisfactory at smaller scales (65.30\% accuracy), it is not suitable as the grid size increases. For example, with a 25x25 lattice, it provides 53\% accuracy. This degradation is attributed to excessive dimensionality reduction: by discarding a substantial portion of the global parity-check matrix, the system loses critical topological information and inter-qubit correlations, ultimately undermining the decoder's decision-making. Consequently, these findings demonstrate that SVD optimization is most effective when employed as a local preprocessing filter rather than as a global compression mechanism.

\subsection{Impact of the Energy Threshold}
A subsequent analysis investigated the correlation between the percentage of information retained through SVD and the resulting logical accuracy of the correction process. Using a $13 \times 13$ physical qubit lattice distributed across sixteen QPUs with a physical error rate $p_{\text{err}}0.01\%$, the SVD retention threshold was systematically varied to assess its impact on decoding performance.

\begin{table}[ht]
    \centering
    \scriptsize
    \caption{Comparison of Logic Accuracy as the threshold energy varies.}
    \label{tab:energy_threshold}
    \begin{tabular}{|l|c|r|}
        \hline
        \textbf{Energy Threshold} & \textbf{Accuracy [\%]}\\ \hline
        1.0            & 73.3  \\ \hline
        0.99            & 78.8   \\ \hline
        0.98            & 76.4   \\ \hline
        0.97           & 77.4     \\ \hline
        0.96            & 79.7    \\ \hline
        0.95            & 76.7     \\ \hline
        
    \end{tabular}
\end{table}

\begin{figure}
    \centering
    \includegraphics[width=0.9\linewidth]{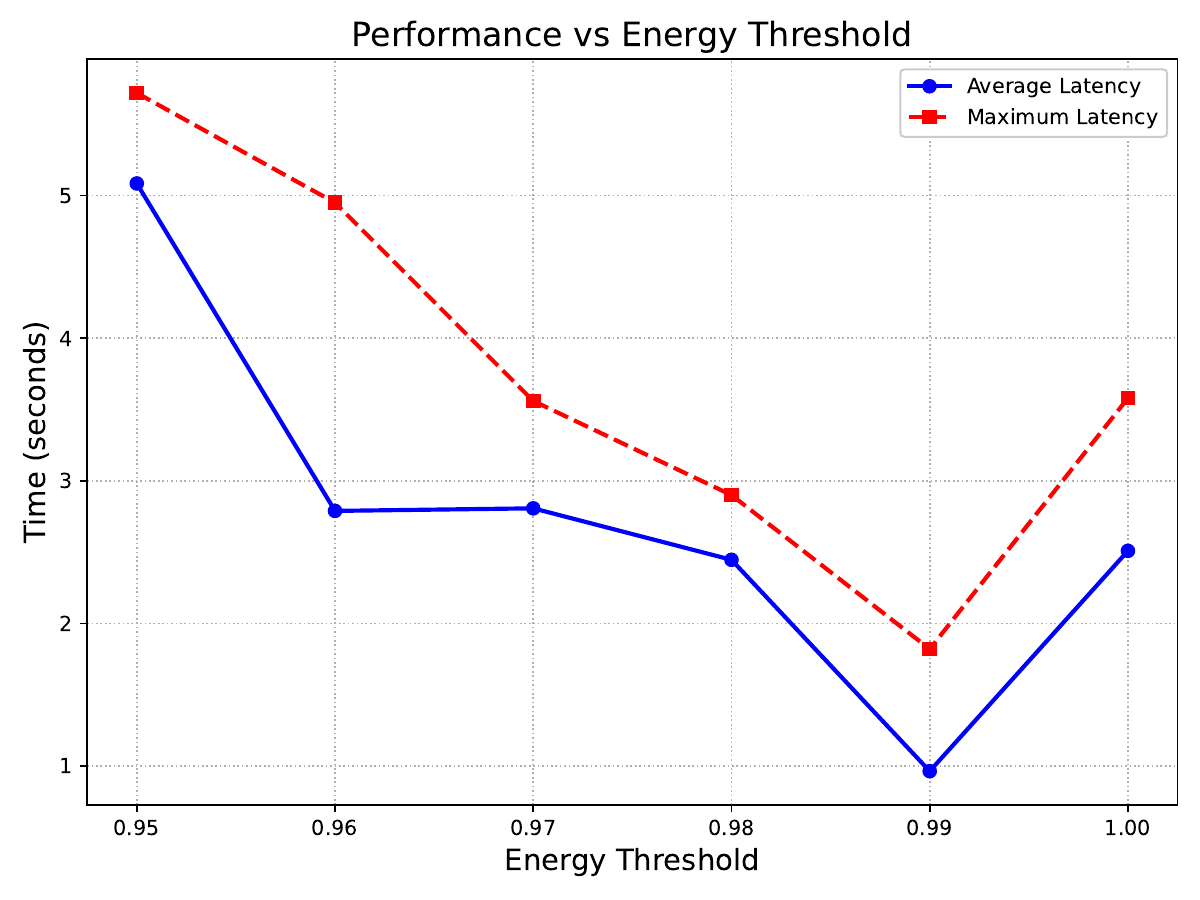}
    \caption{Execution time versus energy threshold.}
    \label{fig:execution_time_energy_threshold}
\end{figure}

As illustrated in Table \ref{tab:energy_threshold}, the logical accuracy exhibits a non-monotonic trend with respect to the SVD energy threshold. Interestingly, the accuracy does not strictly decrease with the threshold; after an initial peak at 78.80\% (threshold 0.99), fluctuations followed by a significant recovery can be observed. This non-monotonic behavior suggests that while aggressive truncation of the $H$ sub-matrices initially restricts the coordinator's decision space, specific compression levels might inadvertently filter out noise or simplify the syndrome landscape in a way that benefits the global OSD decoding process. However, this approach achieves a decoding accuracy on par with the standard OSD reference baseline, demonstrating that the integration of SVD does not introduce significant performance degradation. Furthermore, the execution time generally decreases as the energy threshold is reduced. Minor deviations from this trend are mainly attributable to the stochastic nature of the simulations, as the measured performance is highly dependent on the specific distribution and occurrence of error events, as shown in Fig. \ref{fig:execution_time_energy_threshold}.

\subsection{Scalability}
The final phase of our analysis investigates how logical accuracy varies as a function of the number of QPUs. 
The experimental setup was designed to ensure that the qubit distribution among QPUs is as uniform as possible. To evaluate the robustness of the distributed protocol, two distinct error regimes were considered.

\begin{itemize}
    \item Stress-Test Regime ($p_{\text{err}}=1\%$): Designed to investigate the system's behavior in edge cases near the theoretical fault-tolerance threshold.
    \item Operational Regime ($p_{\text{err}}=0.1\%$): Designed to reflect the error rates characteristic of state-of-the-art quantum hardware.
\end{itemize}

The results in Table \ref{tab:scalability_result} demonstrate that logical accuracy scales positively with the number of QPUs. This trend is directly attributable to the distribution of the SVD approximation error; as the number of nodes increases, each QPU manages smaller sub-matrices, thereby minimizing the information loss occurring at the local level. Consequently, the cumulative impact of local compression on the global decoder's performance is significantly reduced compared to highly compressed, large-scale local matrices.

\begin{table}[ht]
    \centering
    \scriptsize 
    \caption{Comparison of Logic Accuracy versus Number of QPUs}
    \label{tab:scalability_result}
    \begin{tabular}{|c|c|c|c|}
        \hline
        \textbf{\# QPU} & \textbf{$p_{\text{err}} [\%]$} & \textbf{Accuracy[\%]} & \textbf{\# CNOT} \\ \hline
        1        & 1.0    & 73.1                             & 624 \\ \hline
        4        & 1.0    & 74.4                              & 676 \\ \hline
        9        & 1.0    & 73.0                              & 728 \\ \hline
        16       & 1.0     & 76.4                              & 780  \\ \hline
        1        & 0.1    & 96.3                             & 624 \\ \hline
        4        & 0.1    & 97.6                              & 676 \\ \hline
        9        & 0.1    & 94.5                              & 728 \\ \hline
        16       & 0.1     & 97.4                              & 780 \\ \hline
        
    \end{tabular}
\end{table}

\begin{figure}[ht!]
\centering
\begin{subfigure}[b]{0.9\linewidth}
    \centering
    \includegraphics[width=\textwidth]{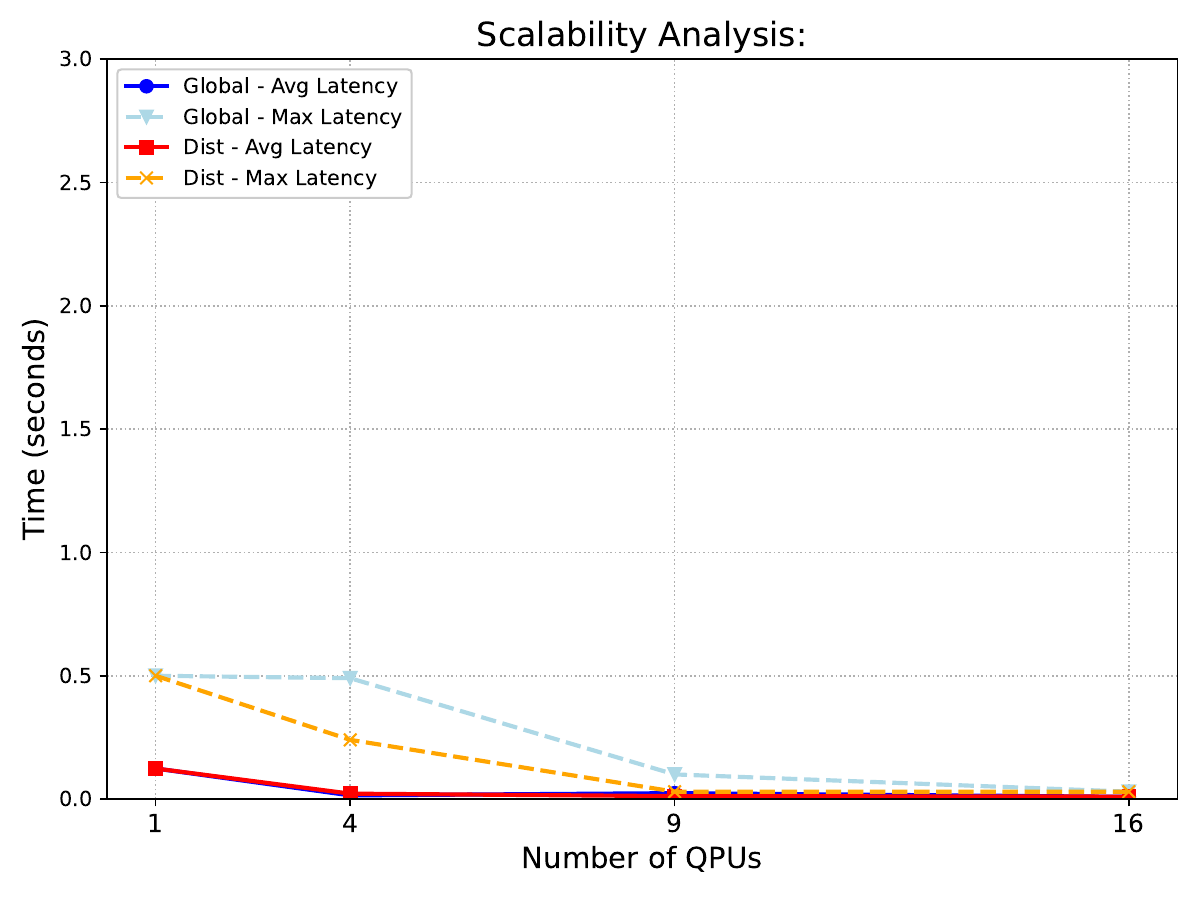}
    \caption{BP-OSD time (real).}
    \label{fig:execution_time_number_nodes1}
\end{subfigure}
\hfill
\begin{subfigure}[b]{0.9\linewidth}
    \centering
    \includegraphics[width=\textwidth]{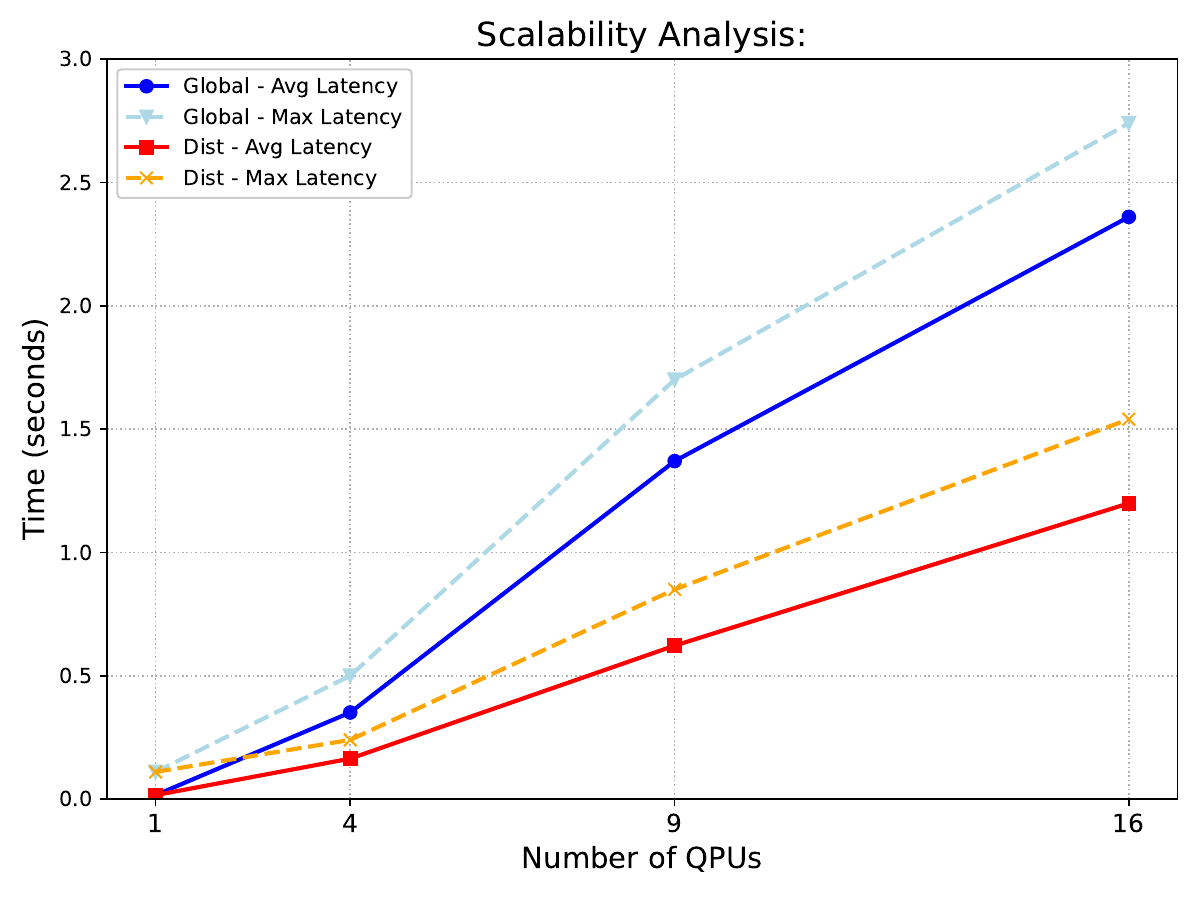}
    \caption{Communication time (simulated).}
    \label{fig:execution_time_number_nodes2}
\end{subfigure}
\caption{Execution time versus number of QPUs.}
\label{fig:fig:execution_time_number_nodes}
\end{figure}

Fig. \ref{fig:execution_time_number_nodes1} illustrates the scaling behavior of the local computational stages (BP-OSD with SVD) as a function of the QPU count. The observed trend closely aligns with the theoretical performance models derived in Section \ref{sec:opt-bposd}. In the same plot, the execution time of the Global SVD described in Section \ref{sec:global} is reported as well. Fig. \ref{fig:execution_time_number_nodes2} demonstrates that the total communication latency scales poorly with the number of nodes. This overhead stems directly from the centralized coordination model: the coordinator must sequentially collect the local, SVD-reduced parity-check matrices from all subgrids and subsequently broadcast the calculated corrections. This serial communication phase represents a  bottleneck that cannot be parallelized.
Nevertheless, two crucial factors must be considered when evaluating these results. First, these benchmarks are obtained via software-based discrete-event simulation, where inter-process communication overhead is significantly more severe than the microsecond-latency exchanges typical of dedicated hardware architectures. Second, for larger topological networks, the computational speedup provided by SVD-driven dimensionality reduction is expected to heavily outweigh this communication penalty, given that the OSD execution time scales cubically with the Matrix dimension.

\section{Conclusion}
\label{sec:conclusion}

A novel parallel QEC decoding approach was introduced and applied to distributed quantum computers. By means of simulations, it was evaluated in terms of complexity, accuracy, and scalability. 

Future work will concern the application of the proposed approach to more advanced surface codes, such as qLDPC ones. Furthermore, it would also be relevant to optimize the BP stage.

\section*{Acknowledgements}
This research benefits from the High Performance Computing facility of the University of Parma, Italy (HPC.unipr.it).

\bibliographystyle{unsrturl}
\bibliography{references}

\end{document}